\documentclass[10pt]{article}

\usepackage[margin=1in]{geometry}
\usepackage{amsmath,amssymb}
\usepackage{microtype}
\usepackage{booktabs}
\usepackage{listings}
\usepackage{xcolor}
\usepackage{enumitem}
\usepackage{cite}
\usepackage{balance}
\usepackage{hyperref}
\usepackage{xurl}

\hypersetup{
  colorlinks=true,
  linkcolor=blue,
  citecolor=blue,
  urlcolor=blue
}

\title{Memelang: An Axial Grammar for LLM-Generated Vector-Relational Queries}
\author{Bri Holt\\\texttt{bri@memelang.net}}
\date{}

\begin{document}

\twocolumn[
\maketitle
\begin{abstract}
Structured generation for LLM tool use highlights the value of compact DSL intermediate representations (IRs) that can be emitted directly and parsed deterministically. This paper introduces \emph{axial grammar}: linear token sequences that recover multi-dimensional structure from the placement of rank-specific separator tokens. A single left-to-right pass assigns each token a coordinate in an $n$-dimensional grid, enabling deterministic parsing without parentheses or clause-heavy surface syntax. This grammar is instantiated in \emph{Memelang}, a compact query language intended as an LLM-emittable IR whose fixed coordinate roles map directly to table/column/value slots. Memelang supports coordinate-stable relative references, parse-time variable binding, and implicit context carry-forward to reduce repetition in LLM-produced queries. It also encodes grouping, aggregation, and ordering via inline tags on value terms, allowing grouped execution plans to be derived in one streaming pass over the coordinate-indexed representation. Provided are a reference lexer/parser and a compiler that emits parameterized PostgreSQL SQL (optionally using pgvector operators).
\end{abstract}
\medskip
]

\section{Introduction}
This paper proposes a compact intermediate representation (IR) designed for direct LLM emission and deterministic compilation into hybrid vector-relational queries. Focus is on minimizing structural degrees of freedom in the emitted string while preserving a straightforward compilation path to parameterized SQL.

\paragraph{Task definition.}
Given a natural-language query and a database schema, an LLM produces an IR string that (i) is compact with a low-token count, (ii) is \emph{deterministically parseable} into a structured representation with well-typed roles, and (iii) can be compiled into a \emph{parameterized} execution plan supporting hybrid vector-relational evaluation.

\paragraph{Design goals.}
\begin{itemize}[itemsep=0.2em, topsep=0.2em]
  \item \textbf{Deterministic parsing without nesting.} The syntax must admit a unique parse without parentheses, indentation sensitivity, or clause-heavy keywords.
  \item \textbf{Streaming-friendly structure.} Parsing should be realizable in a single left-to-right pass, enabling incremental validation and early error detection (and optionally prefix-valid completion under constrained decoding).
  \item \textbf{Low-entropy surface form.} The representation should minimize token count and syntactic variation, reducing the burden on the generator and limiting equivalent-but-different clause permutations.
  \item \textbf{Bounded structural degrees of freedom.} Semantic roles (e.g., table/column/value) should be associated with fixed positions to constrain output space and improve reliability under generation.
  \item \textbf{Safe compilation boundary.} Compilation should preserve parameterization of user-supplied values and avoid string-concatenated query construction, reducing injection risk and enabling backend-side planning and caching.
\end{itemize}

\paragraph{Non-goals.}
Memelang is not intended to be (i) a general-purpose programming language, (ii) a full reimplementation of SQL (including arbitrary nesting, correlated subqueries, or engine-specific procedural features), or (iii) an optimizer specification. The focus is a compact, LLM-emittable IR with deterministic structure that supports common hybrid query patterns and admits predictable compilation and validation.

\paragraph{Contributions.}
This paper makes the following contributions:
\begin{itemize}[itemsep=0.2em, topsep=0.2em]
  \item Defines \emph{axial grammar} and a deterministic single-pass coordinate assignment scheme for separator-ranked token streams.
  \item Presents \emph{Memelang}, an axial-grammar query IR with fixed slot semantics (table/column/value), relative references, parse-time variable binding, and inline tags for grouping/aggregation/ordering.
  \item Provides a reference lexer/parser, a coordinate-indexed IR, and a compiler that emits parameterized PostgreSQL SQL (including optional \texttt{pgvector} similarity operators).
\end{itemize}

\smallskip
Section~\ref{sec:related} situates axial grammars relative to constrained decoding, Text-to-SQL IRs, and hybrid vector–relational systems.

\section{Related Work}
\label{sec:related}

\subsection{Grammar-constrained decoding and structured generation IRs}
A large body of work treats structured LLM output as a \emph{constrained generation} problem, where decoding is restricted to strings accepted by a grammar or schema. Incremental parsing for constrained decoding has been shown effective for structured targets such as SQL, notably via parser-in-the-loop filtering during generation \cite{picard}. More generally, grammar-constrained decoding without finetuning \cite{gcd-without-finetuning} and recent constrained-reasoning and correctness-focused decoding frameworks \cite{crane,correctness-guaranteed-constrained-decoding} emphasize that \emph{representation choice} can reduce search complexity by shrinking the set of syntactically valid continuations. Empirically, structured generation has increasingly standardized around JSON Schema as a constraint language; recent benchmarking work (JSONSchemaBench) compares constrained-decoding frameworks and highlights practical trade-offs in schema coverage, efficiency, and downstream quality \cite{jsonschemabench}. Memelang is complementary to these efforts: rather than proposing a new decoding algorithm, it proposes an \emph{IR surface form} whose structure is induced by ranked separators and fixed coordinate roles, reducing clause permutations and enabling deterministic parsing with minimal delimiter overhead.

\subsection{Text-to-SQL intermediate representations}
Intermediate representations (IRs) are a recurring strategy in Text-to-SQL to bridge the mismatch between natural language intent and SQL clause mechanics. IRNet generates an intermediate SemQL form and deterministically maps it to SQL \cite{irnet}, while NatSQL explicitly removes or normalizes parts of SQL that are difficult to align with natural language (e.g., certain keywords and nesting) to simplify generation \cite{natsql}. Separately, strong Text-to-SQL parsers often rely on structured schema encoding and explicit SQL structure modeling, e.g., relation-aware schema encoding and linking \cite{ratsql} or bottom-up construction of structured programs \cite{smbop}. Recent LLM-era Text-to-SQL systems also emphasize execution feedback and search/correction loops, as well as lightweight frameworks with improved schema linking and self-correction \cite{lite-sql,sql-o1,excot}.
Most cross-domain Text-to-SQL results are still reported on benchmark suites such as Spider \cite{spider}, with continued work on updating legacy datasets for modern LLM usage \cite{llmsql} and on multilingual evaluation settings \cite{multilingual-text2sql}. Memelang shares the same broad motivation of lowering the burden of emitting correct SQL, but differs in the \emph{mechanism}: it fixes table/column/value roles by coordinate position (rather than by SQL clause syntax), and it uses ranked separators to recover structure in a single pass, aiming for a lower-entropy, LLM-emittable IR that remains compile-friendly for hybrid workloads. For broader overviews of LLM-based Text-to-SQL methods and trends, see recent surveys \cite{survey-llm-text2sql-hong,survey-text2sql-era-llms,survey-llm-enhanced-text2sql}.

\subsection{Query languages and linearizations for neural generation}
A common design axis in semantic parsing and program synthesis is how to linearize hierarchical structure for sequence models: action sequences over grammars and AST-driven generation explicitly expose syntactic constraints \cite{yin-neubig-2017}. In Text-to-SQL, structured decoders and AST-like representations similarly reduce invalid programs by tying generation to grammar productions or compositional substructures \cite{ratsql,smbop}. Alternative linearizations (e.g., S-expressions, stack languages, or indentation-sensitive formats) typically retain explicit nesting tokens or rely on whitespace structure. In contrast, \emph{axial grammars} linearize without explicit nesting: multi-dimensional structure is inferred from the placement of a small number of ranked separators, and semantic roles are assigned by stable coordinate indices. This is the primary novelty relative to prior linearizations: Memelang's separators encode axis transitions, while coordinate roles encode semantic slots, jointly constraining the space of valid strings without parentheses-heavy syntax.

\subsection{Hybrid vector-relational query models and systems}
Recent systems work argues for unified representations and optimizers for hybrid workloads that mix relational predicates with vector similarity search \cite{text2vectorsql,chase,exqutor,arcade,hmgi}. Surveys of vector DBMS designs and of filtered ANN (vector+scalar) workloads further emphasize that hybrid queries are now a primary interface requirement, not an edge case \cite{survey-vdbms,survey-fanns}. Beyond query models, recent work also highlights production concerns such as scalable embedding search services \cite{hakes} and software testing/reliability roadmaps for vector DBMS deployments \cite{reliable-vdbms-testing-roadmap}. These contributions primarily target query processing, indexing, and execution planning. Memelang targets a different bottleneck: \emph{the generator-facing query IR} that an LLM can emit reliably while still compiling to a hybrid execution plan. The design goal is to make hybrid constraints (symbolic filters, joins, aggregations, and vector predicates) expressible in a compact, deterministically parseable form, rather than to propose new vector indexing or optimizer techniques. For practical deployments, Memelang's compilation target can align with SQL backends that expose vector operators (e.g., via PostgreSQL extensions such as \texttt{pgvector}) \cite{pgvector}.

\subsection{IRs for tool-using agents and API calling}
LLM tool-use research often frames structured outputs as API calls whose arguments must match a predefined schema. Tool-use can be learned or improved via supervised or self-supervised data that teaches models to invoke tools with correct arguments \cite{toolformer}, and via retrieval-augmented approaches that ground API usage in documentation \cite{gorilla}. Agent prompting approaches such as ReAct interleave reasoning with actions to interact with external systems \cite{react}. In practice, many tool APIs adopt JSON-schema-like specifications for arguments, and commercial function/tool calling interfaces explicitly expose schema-constrained argument generation \cite{openai-function-calling}. Memelang differs from these schema-object approaches by using a \emph{linear, coordinate-typed IR} rather than a nested key-value structure. Its contribution is a compact representation where separator-induced coordinates and fixed slot roles reduce structural ambiguity, while remaining compatible with grammar- or schema-constrained decoding techniques.

\section{Axial Grammar}
An \emph{axial grammar} is a linear token grammar in which a small set of \emph{ranked separators} induce a coordinate system over an otherwise flat stream. Parsing proceeds in one left-to-right scan. Each non-separator token receives a coordinate in an $n$-dimensional lattice. The embodiment then interprets coordinate positions as semantic roles. Grammar maps tokens to coordinates, semantics maps coordinates to roles.

A formal description of axial grammar is given on the next page.

\onecolumn

\paragraph{Tokenization.} Fix an arity $n \in \mathbb{N}_{>0}$. Let $\Sigma$ be a token alphabet partitioned into
ranked separators $S$ and non-separators $A$:
\[
\Sigma = S \;\dot{\cup}\; A.
\]
Each separator has a rank given by a total function
\[
\rho:S \to \{0,1,\dots,n-1\}.
\]

\paragraph{Scan state and coordinate assignment.}
For a token stream $\tau=(t_1,\dots,t_T)\in\Sigma^T$, define the scan state
$\mathbf{i}^{(k)} \in \mathbb{N}^n$ with $\mathbf{i}^{(0)}=\mathbf{0}$ and
\[
\mathbf{i}^{(k)} \;=\;
\begin{cases}
U_{\rho(t_k)}\!\big(\mathbf{i}^{(k-1)}\big) & \text{if } t_k\in S,\\
\mathbf{i}^{(k-1)} & \text{if } t_k\in A,
\end{cases}
\]
where for $\mathbf{i}=(i_{n-1},\dots,i_0)$ and rank $r$,
\[
\big(U_r(\mathbf{i})\big)_m \;=\;
\begin{cases}
i_m & m>r,\\
i_r + 1 & m=r,\\
0 & m<r.
\end{cases}
\]
The (raw) coordinate of each non-separator token is
\[
\phi_\tau(k) \;=\; \mathbf{i}^{(k)} \qquad \text{for all } k \text{ with } t_k\in A.
\]

\paragraph{Cellization (induced grid).}
Define the (raw) cell content map $C_\tau:\mathbb{N}^n\to A^*$ by
\[
C_\tau(\mathbf{x})
\;=\;
\langle\, t_k \;:\; t_k\in A \ \wedge\  \phi_\tau(k)=\mathbf{x}\,\rangle,
\]
preserving stream order. Empty cells have $C_\tau(\mathbf{x})=\epsilon$.

Optionally, an embodiment may apply a (possibly $\tau$-dependent) \emph{bijective} reindexing
\[
\pi_\tau:\mathbb{N}^n \to \mathbb{N}^n,
\]
yielding the effective coordinate $\phi^\pi_\tau(k)=\pi_\tau(\phi_\tau(k))$ and effective cells
\[
C^{\pi}_\tau(\mathbf{x}) = C_\tau\!\big(\pi_\tau^{-1}(\mathbf{x})\big).
\]
For notational convenience we may write $\pi$ when $\tau$ is clear.

\paragraph{Slice-local reversal (reverse indexing).}
To support right-aligned or reversed indexing along rank $r$ within each fixed
higher-rank prefix, allow the reindexing map to be $\tau$-dependent.
For $\mathbf{x}=(x_{n-1},\dots,x_0)$ let the higher-rank prefix be
$\mathbf{x}_{>r}=(x_{n-1},\dots,x_{r+1})$ and define the slice width
\[
W_r^\tau(\mathbf{u})
=
\begin{cases}
0 & \text{if } \nexists\,\mathbf{y}\text{ with } \mathbf{y}_{>r}=\mathbf{u}
\text{ and } C_\tau(\mathbf{y})\neq \epsilon,\\[2pt]
1+\max\{\, y_r : C_\tau(\mathbf{y})\neq \epsilon \wedge \mathbf{y}_{>r}=\mathbf{u}\,\}
& \text{otherwise.}
\end{cases}
\]

Then define the rank-$r$ reversal reindexing $\pi^{\mathrm{rev},r}_\tau:\mathbb{N}^n\to\mathbb{N}^n$ by
\[
\big(\pi^{\mathrm{rev},r}_\tau(\mathbf{x})\big)_m
=
\begin{cases}
x_m & m\neq r,\\[2pt]
W_r^\tau(\mathbf{x}_{>r})-1-x_r & m=r \ \wedge\ x_r < W_r^\tau(\mathbf{x}_{>r}),\\[2pt]
x_r & m=r \ \wedge\ x_r \ge W_r^\tau(\mathbf{x}_{>r}).
\end{cases}
\]
This map is an involution (hence bijective), so $(\pi^{\mathrm{rev},r}_\tau)^{-1}=\pi^{\mathrm{rev},r}_\tau$.

\paragraph{Embodiment interpretation.}
An axial grammar instance is completed by a (possibly partial) cell interpreter
\[
g:A^* \rightharpoonup \mathcal{X},
\]
producing a coordinate-indexed representation (partial map)
\[
E:\mathbb{N}^n \rightharpoonup \mathcal{X},
\qquad
E(\mathbf{x})=
\begin{cases}
g(C^\pi_\tau(\mathbf{x})) & C^\pi_\tau(\mathbf{x})\neq \epsilon,\\
\text{undefined} & \text{otherwise}.
\end{cases}
\]
Write $E(\mathbf{x})\downarrow$ if defined and $E(\mathbf{x})\uparrow$ if undefined.

\paragraph{Carry-forward closure.}
For rank $r$, let $\mathbf{e}_r\in\mathbb{N}^n$ be the $r$-th unit vector.
Define the carry-forward (inheritance) completion $\operatorname{cf}_r(E)$ by
\[
\operatorname{cf}_r(E)(\mathbf{x})=
\begin{cases}
E(\mathbf{x}) & E(\mathbf{x})\downarrow,\\
\operatorname{cf}_r(E)(\mathbf{x}-\mathbf{e}_r)
& E(\mathbf{x})\uparrow \ \wedge\ x_r>0,\\
\bot & \text{otherwise},
\end{cases}
\]
where $\bot$ denotes absence after inheritance.

\paragraph{Relative references.}
Let $R\subseteq A$ be designated reference atoms with offsets
\[
\Delta:R \to \mathbb{Z}^n.
\]
For $\mathbf{x}\in\mathbb{N}^n$ and $a\in R$, define (when $\mathbf{x}+\Delta(a)\in\mathbb{N}^n$)
\[
\operatorname{res}(a,\mathbf{x};E)
\;=\;
\operatorname{cf}_r(E)\big(\mathbf{x}+\Delta(a)\big),
\]
for the embodiment-chosen inheritance rank $r$ (or replace $\operatorname{cf}_r$ by any
chosen normalization operator).

\paragraph{Variable binding (single-pass environment).}
Let $\mathcal{V}$ be a set of variable names. Let $\prec$ order \emph{effective} coordinates by first occurrence:
\[
\mathbf{x}\prec \mathbf{y}
\iff
\min\{k:\, t_k\in A \wedge \phi^\pi_\tau(k)=\mathbf{x}\} < \min\{k:\, t_k\in A \wedge \phi^\pi_\tau(k)=\mathbf{y}\}.
\]
A binding policy is any left-to-right construction of an environment
$\beta:\mathcal{V}\rightharpoonup \mathbb{N}^n$ over $\prec$-ordered cells:
\[
\beta_{\text{new}} \;=\; \beta_{\text{old}} \cup \{\, v \mapsto \mathbf{x} \;:\; v \text{ is bound at } \mathbf{x}\,\}.
\]
A reference to $v$ denotes $E(\beta(v))$ (and is an error if $\beta(v)$ is undefined).

\medskip
This defines axial grammars as (i) a separator-ranked stream transducer $\phi_\tau$ producing an
$n$-dimensional coordinate lattice, plus (ii) an embodiment map $g$ (and optional normalizations such as
$\operatorname{cf}_r$, relative offsets $\Delta$, and bindings $\beta$). For typability, Axis:$r$:$i$ denotes the coordinate index $i_r=i$ on rank $r$.

\twocolumn
\section{Memelang Syntax \& Semantics}
Memelang instantiates an axial grammar with three ranks (Matrix, Vector, Limit) shown in Table~\ref{tab:axisrank} and fixed semantic roles on the Limit axis (Table, Column, Value) index shown in Table~\ref{tab:indexmap}.

\begin{table}[h]
\centering
\caption{Memelang axis rank mapping.}
\label{tab:axisrank}
\begin{tabular}{@{}lll@{}}
\toprule
Rank & Name & Separator \\
\midrule
Axis:2 & Matrix & Double semicolon (\texttt{;;}) \\
Axis:1 & Vector & Single semicolon (\texttt{;}) \\
Axis:0 & Limit  & Whitespace       (\texttt{\textbackslash{}s+}) \\
\bottomrule
\end{tabular}
\end{table}

For Axis:0 separation, one or more whitespace characters are lexed as a single separator token. Leading/trailing whitespace inside vectors is normalized.

Token semantic roles are a function of their recovered coordinates, eliminating reliance on clause keywords, parentheses, or hierarchical scope markers. Within each Vector, Axis:0 indices are right-aligned so that the rightmost Limit occupies Axis:0:0, regardless of vector length. Axis:0:2 is the table slot, Axis:0:1 the column slot, and Axis:0:0 the value slot (Table~\ref{tab:indexmap}).

\begin{table}[h]
\centering
\caption{Memelang Axis:0 index mapping.}
\label{tab:indexmap}
\begin{tabular}{@{}ll@{}}
\toprule
Index & Role \\
\midrule
Axis:0:2 & Table identifier \\
Axis:0:1 & Column identifier \\
Axis:0:0 & Value term / predicate \\
\bottomrule
\end{tabular}
\end{table}

\smallskip
\noindent Memelang's surface syntax uses a small set of reserved tokens to express wildcards, relative references, comparisons, and disjunctions:
\begin{itemize}[itemsep=0.2em, topsep=0.2em]
  \item \textbf{Wildcard.} A lone underscore (\texttt{\_}) denotes a wildcard atom, matching any value in any slot.
  \item \textbf{Vector-same.} An at-sign (\texttt{@}) refers to the resolved value of the same Limit index of the prior Vector of the current Matrix.
  \item \textbf{Matrix-same.} A caret (\texttt{\textasciicircum}) refers to the resolved value of the same Limit index in the last Vector of a previous Matrix.
  \item \textbf{Comparators.} A comparison operator token (\texttt{=}, \texttt{!=}, \texttt{>}, \texttt{>=}, \texttt{<}, \texttt{<=}, \texttt{\textasciitilde}, \texttt{!\textasciitilde}) optionally controls constraint intersection during evaluation; omitting an operator defaults to equality (\texttt{=}).
  \item \textbf{Literals.} Tokens such as ALNUM (\texttt{joe}), QUOT (\texttt{"Joe Smith"}), INT (\texttt{5}), and DEC (\texttt{5.1}) denote literal atoms.
  \item \textbf{Disjunction.} A comma (\texttt{,}) logically disjoins (ORs) alternative values within a single expression.
\end{itemize}

All columns in Axis:0:1 column slot are projected. Memelang permits non-fixed tokens in the Axis:0 table and column slots, allowing a single value predicate to range over multiple columns and/or tables. This is syntactically valid in Memelang but is not broadly supported by contemporary database engines. In practice, a wildcard in the column slot or a vector consisting of only a wildcard in the value slot projects all table columns.

FUNC annotations attach grouping, aggregation, and ordering to the value slot. Syntactically, a FUNC is a colon-prefixed tag chain placed at the start of the Axis:0:0 term (e.g., \texttt{:min:asc>=4.2}). Aggregate tags (\texttt{:sum, :cnt, :min, :max, :avg, :last}) compile to SQL aggregate expressions over the Vector's column; \texttt{:grp} marks a grouping key; \texttt{:asc} and \texttt{:des} specify ordering on the (possibly aggregated) expression.

A binding tag of the form \texttt{:\$x} attaches to the value slot at the point of introduction; later occurrences of \texttt{\$x} in the same Matrix are interpreted as references to the bound coordinate expression. Unbound variables are treated as errors.

\subsection{Memelang query grammar}
A concise EBNF-style fragment for query mode is shown below:

\begin{lstlisting}[language={},caption={Minimal EBNF (query mode).}]
query      := (matrix ";;")+ ;
matrix     := vector (";" vector)* ;
vector     := limit (WS limit)* ;
limit      := left | right | (cmp right) | (left cmp right) ;
left       := (term)? (":" func)* ;
right      := term ("," term)* ;
term       := atom | (mod atom) | (atom mod atom) ;
atom       := ALNUM | QUOT | INT | DEC | "_" | "@" | "$var" | EMB ;
cmp        := "=" | "!=" | "<" | "<=" | ">" | ">=" | "~" | "!~" ;
mod        := "<->" | "<#>" | "<=>" | "+" | "-" | "*" | "/" | "%" | "**" ;
\end{lstlisting}

\section{Memelang Examples}
\label{sec:examples}
\subsection{Scalar filter with carry-forward}

Consider the following Memelang query:

\smallskip
\begin{lstlisting}[language={}]
movies year <1970; title _;;
\end{lstlisting}

\smallskip
This Matrix contains two Vectors separated by \texttt{;}, and terminates with \texttt{;;}. In the first Vector, the three Limit slots (Table, Column, Value) are instantiated as: Table \texttt{movies}, Column \texttt{year}, and a Value predicate \texttt{<1970}. In the second Vector, the Table slot is omitted and is inherited from the prior Vector; the Column slot is \texttt{title}, and the Value slot is the wildcard \texttt{\_}, indicating projection of \texttt{title} with no additional constraint.

\smallskip
\begin{table}[h]
\centering
\caption{Example 1 coordinates as (Axis:2, Axis:1, Axis:0).}
\label{tab:ex1-coords}
\begin{tabular}{@{}lll@{}}
\toprule
Expression & Coordinates & Role \\
\midrule
\texttt{movies}   & (0, 0, 2) & Table \\
\texttt{year}     & (0, 0, 1) & Column \\
\texttt{<1970}    & (0, 0, 0) & Value / predicate \\
\texttt{(@)}      & (0, 1, 2) & Table (carried) \\
\texttt{title}    & (0, 1, 1) & Column \\
\texttt{\_}       & (0, 1, 0) & Value (wildcard) \\
\bottomrule
\end{tabular}
\end{table}

\smallskip
Compiling this IR to PostgreSQL SQL yields:

\smallskip
\begin{lstlisting}[language={}]
SELECT t0.year, t0.title FROM movies AS t0 WHERE t0.year < 1970;
\end{lstlisting}

\subsection{Two-column output with cosine distance and ordering}

The following query returns (i) a cosine-distance expression over \texttt{movies.description} against a text prompt, and (ii) the movie title. The \texttt{:asc} tag orders results by the distance expression in ascending order (closest first).

\smallskip
\begin{lstlisting}[language={}]
movies description <=>"robot":asc<0.35;title _;;
\end{lstlisting}

\smallskip
The first Vector sets Table=\texttt{movies}, Column=\texttt{description}, and the Value slot encodes a cosine-distance expression \texttt{<=>"robot"} with an inline sort tag \texttt{:asc}, filtered by a threshold \texttt{<0.35}. The second Vector projects \texttt{title} and inherits the table slot from the prior Vector.

In the reference implementation, the embedding hook replaces the quoted text with a vector literal; here \texttt{[...]} represents an embedding.

\smallskip
\begin{lstlisting}[language={}]
SELECT (t0.description <=> '[...]'::VECTOR), t0.title
FROM movies AS t0
WHERE (t0.description <=> '[...]'::VECTOR) < 0.35
ORDER BY (t0.description <=> '[...]'::VECTOR) ASC;
\end{lstlisting}

\subsection{Co-star self-join with binding}

A common co-star query can be expressed as a self-join on \texttt{roles.movie}, anchored by a bound
actor name. The following Memelang query binds the anchor name to \texttt{\$a}, joins a second
instance of \texttt{roles} on the movie column via \texttt{@ @ @}, and excludes the anchor actor via
\texttt{!=\$a}.

\smallskip
\begin{lstlisting}[language={}]
roles actor :$a="Bruce Willis";movie _;
@ @ @;actor !=$a;;
\end{lstlisting}

\smallskip
Here, the first Vector constrains \texttt{roles.actor} and binds the introduced value to \texttt{\$a}.
The second Vector projects \texttt{roles.movie}. The third Vector \texttt{@ @ @} introduces a new
instance of the prior table and equates its \texttt{movie} column to the prior Vector's
\texttt{movie} value. The \texttt{@} in the table slot triggers a self-join. The final Vector projects the co-star actor name from the joined instance while
excluding \texttt{\$a}.

An SQL rendering is:

\smallskip
\begin{lstlisting}[language={}]
SELECT t0.actor,
       t0.movie,
       t1.movie,
       t1.actor
FROM roles AS t0, roles AS t1
WHERE t0.actor = 'Bruce Willis'
  AND t1.movie = t0.movie
  AND t1.actor != t0.actor;
\end{lstlisting}

\subsection{Grouped join with vector predicate and meta limit}

The following query targets ``war''-themed movies released before 1980 and returns the top 12 titles
ranked by the \emph{minimum} role rating observed for that movie:

\begin{lstlisting}[language={}]
% War stories before 1980: top 12 movies by minimum role rating
movies year <1980;description <=>"war"<=\$sim;title :grp;roles movie @;rating :min:des;%m lim 12;;
\end{lstlisting}

This Matrix contains five query-mode Vectors followed by one meta-mode Vector:

\begin{itemize}[itemsep=0.2em, topsep=0.2em]
  \item \textbf{Vector 1 (\texttt{movies year <1980}).} Introduces table \texttt{movies} and filters by \texttt{movies.year < 1980}.
  \item \textbf{Vector 2 (\texttt{description <=>"war"<=\$sim}).} Carries forward the table slot (\texttt{movies}), sets the column to
        \texttt{description}, and constrains the inline cosine-distance expression against the embedded prompt
        \texttt{"war"} relative to the runtime parameter \texttt{\$sim}.
  \item \textbf{Vector 3 (\texttt{title :grp}).} Projects \texttt{movies.title} and marks it as the grouping key via \texttt{:grp}.
  \item \textbf{Vector 4 (\texttt{roles movie @}).} Introduces a second table instance \texttt{roles} and joins by setting
        \texttt{roles.movie} equal to \texttt{@}, i.e., the prior Vector's value slot (here, the grouped \texttt{movies.title}).
  \item \textbf{Vector 5 (\texttt{rating :min:des}).} Carries forward \texttt{roles}, selects the aggregate \texttt{:min} over
        \texttt{roles.rating}, and orders by this aggregate descending via \texttt{:des}.
  \item \textbf{Meta (\texttt{\%m lim 12}).} Sets a compilation-time limit (\texttt{LIMIT 12}) without introducing additional query predicates.
\end{itemize}

In the reference compiler, grouped queries require every projected expression to be either a grouping key (\texttt{:grp})
or an aggregate. Expressions not explicitly tagged are assigned a default aggregate (\texttt{:last}) to satisfy SQL's
\texttt{GROUP BY} requirements; this policy does not change the earlier filter predicates.

Compiling to PostgreSQL SQL yields:

\begin{lstlisting}[language={}]
SELECT
  MAX(t0.year),
  MAX(t0.description <=> [...]),
  t0.title,
  MAX(t1.movie),
  MIN(t1.rating)
FROM movies AS t0, roles AS t1
WHERE t0.year < 1980
  AND (t0.description <=> [...]) <= $sim
  AND t1.movie = t0.title
GROUP BY t0.title
ORDER BY MIN(t1.rating) DESC
LIMIT 12;
\end{lstlisting}

Here the embedding hook supplies the vector parameters for \texttt{"war"}, and \texttt{\$sim} is compiled as a parameter provided by the runtime (or overridden via meta directives).

\section{Conclusion}
Memelang proposes an axial grammar that serves as an LLM-oriented, neuro-symbolic intermediate representation for unified vector-relational querying. The core design---a tensor-shaped AST with carry-forward and binding---reduces redundancy, improves deterministic parseability, and yields a clean compilation path to parameterized SQL for PostgreSQL. By integrating symbolic constraints (tables, columns, joins, aggregates) with dense vector semantics (distance operators and embedding hooks), Memelang targets a practical gap in current LLM-to-database workflows: compactness and correctness under constrained generation.

\noindent\textbf{Website:} \url{https://www.memelang.net/}

\smallskip
\noindent\textbf{Repository:} \url{https://github.com/memelang-net/memesql9}

\smallskip
\noindent\textbf{Patent:} U.S. Patent 12,475,098 with additional patents pending.

\smallskip

\balance

\end{document}